\documentclass[preprint]{JHEP3} 

\usepackage{epsfig,multicol}

\newcommand\fverb{\setbox\pippobox=\hbox\bgroup\verb}
\newcommand\fverbdo{\egroup\medskip\noindent%
            \fbox{\unhbox\pippobox}\ }
\newcommand\fverbit{\egroup\item[\fbox{\unhbox\pippobox}]}
\newbox\pippobox

\newcommand{\be}{\begin{equation}}
\newcommand{\ee}{\end{equation}}
\newcommand{\ba}{\begin{eqnarray}}
\newcommand{\ea}{\end{eqnarray}}
\newcommand{\bt}{\begin{table}}
\newcommand{\et}{\end{table}}
\newcommand{\brt}{\begin{ruledtabular}}
\newcommand{\ert}{\end{ruledtabular}}
\newcommand{\btu}{\begin{tabular}}
\newcommand{\etu}{\end{tabular}}

\def\ov{\over}

\title{Systematics of One-Loop Scattering Amplitudes in $N=4$ Super Yang-Mills Theories}

\author{Mingxing Luo and Congkao Wen\\
Zhejiang Institute of Modern Physics, Department of Physics \\
Zhejiang University, Hangzhou, Zhejiang 310027, P R China \\
    E-mail: \email{luo@zimp.zju.edu.cn, wenagua@sina.com}}
\date{\today}        

\preprint{\hepth{0410118}}  

\abstract{One-loop scattering amplitudes in $N=4$ super Yang-Mills (SYM) theories are analyzed
in the paradigm of maximal helicity violating Feynman diagrams.
There are very limited number of loop integrals to be evaluated. 
For a process with $n$ external particles, there are only $[n/2]-1$ generically independent integrals.
Furthermore, the relations between leading $N_c$ amplitudes $A_{n;1}$ and sub-leading amplitudes $A_{n;c}$ 
are found to be identical to those obtained from conventional field theory calculations,
which can be interpreted as an indirect support for the paradigm.}


\begin{document}


\section{Introduction}

In perturbative Yang-Mills theories,
the Parke-Taylor formula for maximal helicity violating (MHV) amplitudes summarizes much of its elegance \cite{PT}. 
For theories including supersymmetries, 
some of the salient features persist beyond the tree level \cite{bern1,bern2}.
Over the years, various techniques were developed to elucidate these properties.
Among them, the color decomposition method and the spinor helicity technique \cite{xzc} have been
proved to be extremely useful and efficient \cite{rvs}.

Recently, Witten pointed out a deep relation \cite{witten} between $N=4$ SYM
theories and one type B topological string theory, by re-expressing
SYM scattering amplitudes in the language of twistor theories \cite{penrose}.
Taking advantage of insights thus gained 
and by a careful analysis of known helicity amplitudes, 
Cachazo, Svrcek, and Witten (CSW) \cite{csw} proposed a novel
prescription to calculate tree level amplitudes, 
which uses the MHV amplitudes as vertices to construct all other amplitudes,
thus initiated the paradigm of MHV Feynman diagrams. 
The efficiency of the method is phenomenal and the validity of the method has
been checked by various tree level calculations \cite{wuzhu,khoze,ggk,kosower,wuzhu2,bbk,wuzhu3,gk,zhu}. 
In \cite{bst}, the method was extended to the calculation of one-loop MHV amplitudes
and results from conventional quantum field theory were reproduced in the large $N_c$ limit.
It turns out that the paradigm is independent of the large $N_c$ approximation,
at least in the case of MHV scattering amplitudes to one-loop \cite{lw}. 
The twistor-space structure of one-loop amplitudes are further studied in \cite{csw2,csw3,bern10,cachazo10}.
On the other hand, tree-level
amplitudes were also obtained from connected curves in twistor string theories \cite{rsv1,rsv2,rsv3}.

In this paper, one-loop MHV Feynman diagrams will be classified.
It turns out there are only very limited number of integrals to be evaluated,
though there are numerous loop diagrams. 
For a process with $n$ external particles,
there are only $[n/2]-1$ generically independent integrals.
These integrals depend crucially on the number of external particles of negative helicity, 
but marginally on $n$.
All loop integrals can be obtained from these generic integrals by
appropriately substituting relevant helicities and momenta.
We then prove that the relation between leading $N_c$ amplitudes $A_{n;1}$
and sub-leading amplitudes $A_{n;c}$ 
are identical to those obtained from conventional field theory calculations \cite{bern1}. 

As there are scant one-loop calculations beyond $q=2$ in the paradigm, 
these results are speculative to certain extent.
But given their regularity, these results might merit presentation. 
Given the strong evidences gained so far, 
it is hard to imagine the paradigm would go astray, at least at the one-loop level.
Actually, the relation between $A_{n;1}$ and $A_{n;c}$ as revealed
could be interpreted as an indirect support for the paradigm.

The paper will be organized as the follows. 
In section 2, we review existing one-loop results 
and give a brief introduction to the paradigm of MHV Feynman diagrams. 
In section 3, we give a classification of one-loop MHV Feynman diagrams.
In section 4, we establish the relation between leading $N_c$ amplitudes $A_{n;1}$
and sub-leading amplitudes $A_{n;c}$.
We conclude in section 5.

\section{Review of existing results and the paradigm of MHV Feynman diagrams}
In four dimensional space-time, a momentum $k_\mu$ can be
expressed as a bispinor $k_{a\dot{a}}=k_\mu \sigma^\mu_{a\dot{a}}$. 
For massless particle, $k^2=0$, 
the momentum can be factorized $k_{a\dot{a}}=\lambda_a \tilde{\lambda}_{\dot{a}}$ 
in terms of spinors $\lambda_a$, $\tilde{\lambda}_{\dot{a}}$ 
of positive and negative chirality. 
Spinor products are defined to be
$\langle\lambda_1,\lambda_2\rangle=\epsilon_{ab}\lambda^a_1\lambda^b_2$
and $\langle\tilde{\lambda}_1,\tilde{\lambda}_2\rangle=
\epsilon_{\dot{a}\dot{b}}\tilde{\lambda}^{\dot{a}}_1\tilde{\lambda}^{\dot{b}}_2$,
which are usually abbreviated as $\langle 1,2 \rangle$ and
$[1,2]$.

At tree level, the scattering amplitudes of $n$ gluons with one or
none opposite type of helicity vanish. The amplitudes with two
negative helicity are called maximally helicity violating (MHV)
amplitudes. For $N=4$ SYM theories, a MHV amplitude is given
by the generalized Parke-Taylor formula which includes particles
of all helicity \cite{nair}: 
\be 
i (2\pi)^4 \delta^{(4)} \left( \sum_{i=1}^n \lambda_i \tilde{\lambda}_i \right) 
                 \delta^{(8)} \left( \sum_{i=1}^n \lambda_i \eta^i \right) 
                  A_n \left(\{k_i,\lambda_i,a_i\} \right) 
\ee 
where  $\eta_A^i$ are
anti-commuting variables, $A$ is an index of the anti-fundamental
representation of $SU(4)$; $k_i$, $\lambda_i$, and $a_i$ are the
momentum, helicity, and the color index of the $i$-th external
particles, respectively; 
\be A_n \left( \{k_i,\lambda_i,a_i\} \right) = 
      \sum_{\sigma \in S_n/Z_n} {\rm Tr} \left(T^{a_{\sigma(1)}} \cdots T^{a_{\sigma(n)}} \right) 
      \prod_{i=1}^n {1 \ov \langle \sigma(i),\sigma(i+1)\rangle}
\ee 
where $S_n/Z_n$ is the set of non-cyclic permutation of
$\{1,\cdots, n\}$. The $U(N_c)$ generators $T^a$ are the set of
hermitian $N_c\times N_c$ matrices normalized such that ${\rm Tr}
(T^a T^b) = \delta^{ab}$.
Here and after the gauge coupling constant $g$ is not included, 
but can be easily recovered when needed. 
The supersymmetric amplitudes can be expended in powers of the $\eta_A^i$, 
and each term of this expression corresponds to a particular scattering amplitude.

To one-loop, the results take the form \cite{bern1}
\be 
A_n^{\rm 1-loop} \left( \{k_i,\lambda_i,a_i\} \right)
  = \sum_{c=1}^{[n/2]+1} \sum_{\sigma \in S_n/S_{n;c}} Gr_{n;c}(\sigma) A_{n;c}(\sigma) 
\label{eq2.3}
\ee 
where $S_n$ is the set of all permutation of $n$ objects 
and $S_{n;c}$ is the subset leaving $Gr_{n;c}$ invariant. 
The color factors are
\ba 
Gr_{n;1} &=& N_c {\rm Tr} \left( T^{a_1} \cdots T^{a_n} \right) \nonumber \\
Gr_{n;c} &=& {\rm Tr} \left( T^{a_1} \cdots T^{a_{c-1}} \right)
             {\rm Tr} \left( T^{a_c} \cdots T^{a_n} \right) 
\ea 
$A_{n;c>1}$ can be obtained by performing appropriate sums over permutations of $A_{n;1}$: 
\be 
A_{n;c}(1,\cdots, c-1; c,\cdots n) = (-)^{c-1} \sum_{\sigma \in COP \{\alpha\}\{\beta\}} A_{n;1}(\sigma) 
\label{Anc_field}
\ee
where $\alpha_i \in \{\alpha\} \equiv \{c-1,\cdots,1\}$,
 $\beta_i \in \{\beta\} \equiv \{c,c+1,\cdots,n\}$ and $COP
 \{\alpha\}\{\beta\}$ is the set of all permutations of
 $\{1,\cdots,n\}$ with $n$ held fixed that preserve the cyclic
 ordering of the $\alpha_i$ within $\{\alpha\}$ and of the $\beta_i$ within
 $\{\beta\}$, while allowing for all possible relative ordering
 of the $\alpha_i$ with respect to the $\beta_i$.

In \cite{csw}, it was proposed that these MHV amplitudes can be
continued to off-shell and be used as vertices to generate tree diagrams for all non-MHV amplitudes.
These new diagrams are dubbed as MHV diagrams.
Including all relevant factors, the $n$-points MHV vertex is 
\be 
\mathcal{V}_n = i {\rm Tr} \left( T^{a_1} \cdots T^{a_n} \right)(2\pi)^4 
        \delta^{(4)} \left( \sum_{i=1}^n \lambda_i \tilde{\lambda}_i \right) 
        \delta^{(8)} \left( \sum_{i=1}^n \lambda_i \eta^i \right)
        \prod_{i=1}^n {1 \ov \langle i,i+1 \rangle} 
\label{eqvn}
\ee 
The propagator for internal lines of momentum $k$ is simply taken as $i/k^2$. The
issue is how to define the corresponding spinor for particles with
off-shell momentum $k$. The CSW prescription is to take an
arbitrary spinor of positive helicity $\tilde{\xi}$ and define $\lambda_a
= k_{a \dot{a}} \tilde{\xi}^{\dot{a}}.$
These rules are extended for applications in one-loop \cite{bst}.
To the leading order of $N_c$, $A_{n;1}$ is consistent 
with the result of conventional quantum field was reproduced in \cite{bst}.
One-loop sub-leading results were reproduced in \cite{lw}.

\section{Classification of MHV Feynman diagrams}
A MHV vertex can be represented by a dot with certain number of lines attached.
Each line represents an external or internal particle 
and there can only be two lines with negative helicity on each vertex.
It is convenient to generate quiver diagrams by linking certain number of dots together. 
The links represent internal particles.
MHV Feynman diagrams are generated by distributing external lines on the dots according to color and helicity
structures. 
Parallel to conventional quantum field theory,
we will call a MHV diagram as one-particle-irreducible (1PI) if the diagram cannot be broken into two pieces 
by cutting one of its internal lines, as one-particle reducible (1PR) if otherwise.
Note there is a difference between 1PI MHV Feynman diagrams and 1PI conventional Feynman diagrams.
1PI MHV Feynman diagrams usually do not correspond to 1PI conventional Feynman diagrams. 

\begin{figure}[h]
\begin{center}
\leavevmode
{\epsfxsize=3.5truein \epsfbox{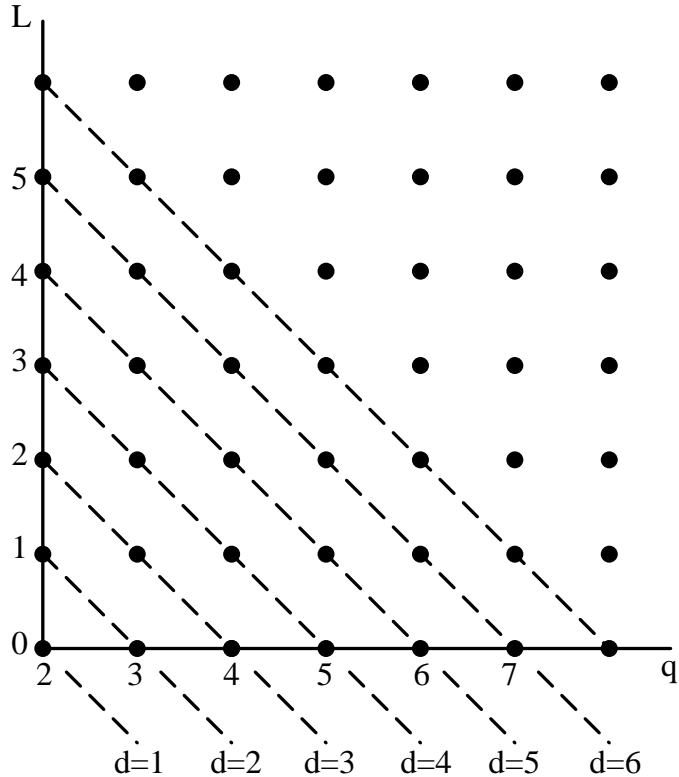}}
\end{center}
\caption{Graphic representation of Eq. (3.1), where $d$ is the degree of the algebraic curve, 
$q$ the number of external particles of negative helicity, and $l$ the number of loops.
Only the lowest line $l=0$ and the point $(q,l)=(2,1)$ are relatively well understood.}
\end{figure}

It was conjectured in \cite{witten} that scattering amplitudes
in $N=4$ SYM theories are supported on certain algebraic curves, if expressed in the twistor-space.
For a process with $q$ particles of negative helicity, the degree of the corresponding algebraic curve is
\be
d=q-1+l \label{eq3.1}
\ee
where $l$ is the number of loops.

Note that $d$ is independent of the total number of external particles in consideration.
In the paradigm of MHV Feynman diagrams, $d$ is the number of MHV vertices in a given diagram.
Graphically expressed (Figure 1),
Eq (\ref{eq3.1}) serves as a good organization principle in the paradigm.
The horizontal line $l=0$ represents all tree diagrams, one point for a specific $q$.
The vertical line $q=2$ represents all loop diagrams for the MHV scattering amplitudes, one point for a specific $l$.
For each point on line $l=0$, 
the corresponding quiver diagrams are generated 
by making different connected trees of $d$ points in all possible manner.
From which one can write out the amplitude in a straightforward manner, 
though it takes efforts to convert them to familiar forms.
Beyond this line, only the point $(q,l)=(2,1)$ has been worked out analytically.

\begin{figure}[h]
\begin{center}
\leavevmode
{\epsfxsize=4.0truein \epsfbox{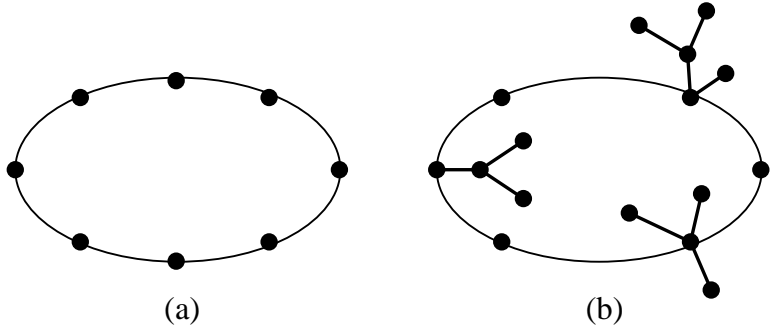}}
\end{center}
\caption{(a) One loop quiver diagrams: (a) 1PI; (b) 1PR.}
\end{figure}

We will call MHV diagrams of $d$ vertices as degree $d$ MHV diagrams.
At one-loop, things simplifies significantly.
$d$ is then equal to $q$, the number of external particles with negative helicity.
And degree $d$ 1PI MHV quiver diagram is a circle with $d$-dots on it (Figure 2a).
The lowest degree $d$ is two, corresponding to the MHV amplitudes, which is also irreducible.
In general, degree $d$ one-loop MHV diagrams is reducible. 
It includes one degree $d$ 1PI MHV diagrams and 
1PR MHV diagrams made of degree $k=\{2,\cdots, d-1\}$ 1PI MHV diagrams by attaching 
trees of total degree $d-k$ in all topologically inequivalent ways (Figure 2b).

MHV Feynman diagrams can also be classified according to their color structures, 
which is closely related to their topologies.
One-loop MHV diagrams with all external lines outside of the circle
will be referred to as leading $N_c$ MHV diagrams, 
since they give the $G_{n;1}$ color factors;
those with external lines both outside and inside the circle
as sub-leading MHV diagrams, 
since they give the $G_{n;c>1}$ color factors. 
(One-loop MHV diagrams with all external lines inside of the circle
are identical to leading MHV diagrams, up to possible overall signs.)
In the next section, it will be proved that all sub-leading diagrams can be expressed 
as sums of the leading ones.
Thus if the loops are calculated iteratively in $d$, 
one encounters one generically new quiver loop diagram for each added degree.
For a process with $n$ external particles, there are $n-3$ generically different quiver loop diagrams in total. 

For a given color factor, one gets all MHV diagrams from relevant quiver diagrams, by distributing
external particles on the vertices in all possible manners
but keeping the relative order according to the color indices.
Specifically, for a process with $n$ external particles and with a degree of $d$,
we first put $m_k$ external lines on the $k$-th vertex, 
$\sum_{k=1}^d m_k = n$.
For each set $\{ m_k \}$, we put on the external lines clock-wise according to the color indices to get a diagram,
and use cyclic permutation to get all different ones:
\be
A_{n}  = \sum_{\{m_k\}} \sum_{\sigma \in (\{\alpha\})} 
\int \prod_{i=1}^d {d^4 L_i \ov L_i^2} d^4 \eta_i
\prod_{k=1}^d \mathcal{V}_{m_k+2}(\sigma)
\ee
where $ \mathcal{V}_{m_k+2}$ is the $m_k+2$-point vertex function given in Eq (\ref{eqvn}),
$L_i$ is the $i$-th internal line, $\eta_i$ the related fermionic variable,
and $(\{\alpha\})$ is the set of all cyclic permutation of $n$ objects.
Obviously, for a given degree $d$, the 1PR integrals can be obtained from those 1PI integrals of degrees less than $d$.
For 1PI cases of degree $d$, 
there are numerous integrals which are generated by distributing external lines on the quiver diagram in all possible ways.
It turns out that all relevant loop integrals can be obtained from one generic integral by
appropriately substituting relevant helicities and momenta.
These integrals depend on $d$ crucially but only marginally on $n$.
More so, for $q>[(n+1)/2]$, the result can be obtained from the case of $q^{'}=n-q$ via parity inversion.
Thus only $[n/2]-1$ of the integrals need to be evaluated eventually.
This trend seems to persist to higher loops, reflecting an intrinsic property implicated by Eq (\ref{eq3.1}).
We note in passing that the leading $N_c$ term of MHV scattering amplitudes at all orders seem to retain their tree level 
helicity structures, provided that the paradigm of MHV Feynman diagrams is also valid to all orders.
There are also regularities in sub-leading orders which need to be further studied \cite{clw}.

\section{The relation between $A_{n;1}$ and $A_{n;c}$}

In this section we will prove that $A_{n;1}$ and $A_{n;c}$ in the paradigm of MHV Feynman diagrams
are related in the same way as those in
conventional quantum field theory. The first part is for 1PI MHV diagrams, 
which is a generalization of the proof in section 4 of \cite{lw}
and includes MHV scattering amplitudes as special cases of $d=2$.
The second part deals with 1PR diagrams.

\subsection{The case of 1PI MHV Feynman diagrams}
One-loop 1PI MHV Feynman diagrams are generated from quiver diagrams as the one in Figure 2a.
We first give the prescription to obtain the leading contribution.
For a given color factor, ${\rm Tr} \left(T^{a_1} \cdots T^{a_n} \right)$,
one gets 
\ba
A_{n;1}^{\rm 1PI} & = & \sum_{\{m_k\}} \sum_{\sigma \in (\{\alpha\})} 
\int \prod_{k=1}^d {d^4 L_k \ov L_k^2} d^4\eta_k
\delta^{(4)} (L_k-L_{k-1}+P_k) \delta^{(8)} (\Theta_k ) F(\{m_k\},\sigma)
\ea
where 
$P_k$ is the total momentum of all external particles on the $k$-th MHV vertex,
$L_k$ is the internal momentum between the $k$-th and $(k+1)$-th vertices,
$\eta_k$ is the related fermionic variables,
$\Theta_k = l_k \eta_k - l_{k-1} \eta_{k-1} + \sum_{i_k = c_k+1}^{c_{k+1}} \lambda_{i_k} \eta_{i_k}$,
$c_k=\sum_{p=1}^{k-1} m_p$, and
\be
F(\{m_k\},\sigma) =
 \prod_{k=1}^d {1 \ov \langle l_k, l_{k-1} \rangle  \langle l_{k-1},\sigma(c_k+1) \rangle 
           \langle \sigma(c_{k+1}), l_k \rangle}
      \prod_{i_k=c_k+1}^{c_{k+1}-1} {1 \ov \langle \sigma(i_k), \sigma(i_k+1)\rangle}.
\ee

For a specific sub-leading term with the color factor
\footnote{Differing from \cite{lw}, we here reverse the order of indices in the first trace.}
\be
{\rm Tr} \left( T^{a_{c-1}} \cdots T^{a_1} \right)
{\rm Tr} \left( T^{a_c} \cdots T^{a_n} \right)
\ee
we construct the sub-leading term $A_{n;c}(c-1,\cdots,1;c,\cdots, n)$
according to Eq (\ref{Anc_field}). A lengthy but straightforward calculation gives
\ba
A_{n;c}^{\rm 1PI}  = (-1)^{c-1} \sum_{\{m_k\}} \sum_{\sigma \in COP^{''} \{\alpha\} \{\beta\}}
\int  \prod_{k=1}^d {d^4 L_k \ov L_k^2} d^4\eta_k  \nonumber \\
\times \delta^{(4)} (L_k-L_{k-1}+P_k) \delta^{(8)} (\Theta_k ) F(\{m_k\},\sigma)
\ea
where $\alpha_i \in \{\alpha\} \equiv \{1,\cdots,c-1\}$,
$\beta_i \in \{\beta\} \equiv \{c,c+1,\cdots,n\}$ 
and $COP^{''} \{\alpha\}\{\beta\}$ is the set of all permutations of
$\{1,\cdots,n\}$ that preserve the cyclic ordering of 
$\alpha_i$ within $\{\alpha\}$ and of $\beta_i$ within $\{\beta\}$, 
while allowing for all possible relative ordering of $\alpha_i$ with respect to $\beta_i$.

In the rest of the section, we will show that 
this $A_{n;c}(c-1,\cdots,1;c,\cdots, n)$ can also be obtained from all relevant sub-leading MHV diagrams. 
Explicitly, these sub-leading MHV diagrams are generated by distributing the first $c-1$ external particles 
inside the circle, and the rest $n-c+1$ external particles outside the circle,
but keeping the cyclic order inside and outside of the circle, respectively.

We now take a specific diagram with $m_k$ external lines on the $k$-th vertex,
among them, $j_k$ of which inside the circle and $i_k$ of which outside of the circle, 
$m_k=i_k+j_k$, as shown on the left of Figure 3.
Denote $I_k=\sum_{p=1}^{k-1} i_p$, $J_k=\sum_{p=1}^{k-1} j_p$.
Applying Eq (4.6) in \cite{lw}, we can make the following transformation for graphs on the left side
of Figure 3 to its right side:
\ba
\prod_{k=1}^d 
\left[ 
\left( \prod_{r=c+I_k+1}^{c+I_{k+1}-1} \frac1{\langle \beta_r,\beta_{r+1} \rangle} \right)
 {1 \ov \langle \beta_{c+I_{k+1}},l_k \rangle \langle l_k,\alpha_{J_{k+1}} \rangle}
\right. \nonumber \\ \left.
\times \left( \prod_{r=J_{k+1}-1}^{J_k+1} \frac1{\langle \alpha_{r+1},\alpha_r\rangle} \right)
 \frac1{\langle \alpha_{J_k+1},l_{k-1}\rangle \langle l_{k-1},\beta_{c+I_k+1} \rangle} 
\right] \nonumber \\
= 
(-1)^{c-1} \prod_{k=1}^d
\left[ {1 \ov \langle \l_k,l_{k-1} \rangle} 
\sum_{\sigma \in COP^{'} \{\alpha^{(k)}\}\{\beta^{(k)}\}} 
{1 \ov \langle l_{k-1},\sigma(c_k+1) \rangle}
\right. \nonumber \\ \left.
\times \left(  \prod_{i_k=c_k+1}^{c_{k+1}-1} {1 \ov  \langle \sigma(i_k),\sigma(i_k+1) \rangle} \right)
{1 \ov \langle \sigma(c_{k+1}), l_k \rangle} \right]
\ea
where $\alpha_{i_k} \in \{\alpha^{(k)}\} \equiv \{J_k+1,\cdots,J_{k+1}\}$, 
 $\beta_{i_k} \in \{\beta^{(k)}\} \equiv \{c+I_k,\cdots,c+I_{k+1}\}$,
$COP^{'}  \{\alpha^{(k)}\}\{\beta^{(k)}\}$  
is the set of all permutations of  $\{1,\cdots,m_k\}$ that preserve the ordering of 
$\alpha_{i_k}$ within $\{\alpha^{(k)}\}$ and of $\beta_{i_k}$ within $\{\beta^{(k)}\}$, 
while allowing for all possible relative ordering of $\alpha_{i_k}$ with respect to $\beta_{i_k}$.
Cyclically permuting all elements in $\{\alpha\}=\prod_{k=1}^d \bigcup \{\alpha^{(k)}\}$ 
and $\{\beta\}=\prod_{k=1}^d \bigcup \{\beta^{(k)}\}$, 
take the sum for all cases of $i_k$, $j_k$ and $m_k$, and including all the $\delta$ functions, we get
\be
\sum_m (-1)^{c-1} \sum_{\sigma \in COP^{''}\{\alpha\} \{\beta\}}
\delta^{(4)} (L_k-L_{k-1}+P_k) \delta^{(8)} (\Theta_k ) F(\{m_k\},\sigma)
\ee 
which is nothing but the integrand in Eq (4.4).

Thus for 1PI MHV Feynman diagrams, we have proved that sub-leading amplitudes $A_{n;c}$ 
obtained from the sub-leading MHV diagrams are related to the leading $N_c$ amplitudes $A_{n;1}$ 
in the same way as those obtained from conventional field theory calculations.
In particular, this relation does not assume any of the external particles to be on-shell, 
since the Schouten identity and our reasoning have not put any requirement on the spinors.
\begin{figure}[h]
\begin{center}
\leavevmode
{\epsfxsize=5.0truein \epsfbox{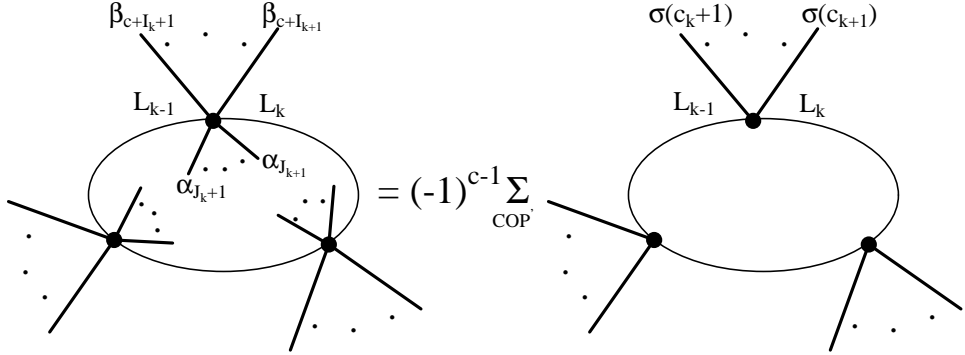}}
\end{center}
\caption{Graphic representation of Eq (4.5), 
a one-loop sub-leading MHV diagram is expressed in terms of a set of one-loop leading MHV diagrams.
All the external lines inside the circle are reflected to outside, in the manner prescribed in 
Eq (4.6) and Figure 7 of [19].}
\end{figure}

\subsection{The case of 1PR MHV Feynman diagrams}
One encounters two subtleties in cases of IPR MHV diagrams.
The first subtlety is that there could be trees both inside and outside the circle.
An inside tree can be reflected to the outside, in almost the same manner as a single external line,
with two modifications:
the order of external lines on the inside the tree has to be kept as if it is a mirror reflection,
and the overall sign is $(-1)^{n_t}$ rather than $-1$, where $n_t$ is the total number of external lines on the tree.
Each tree (inside and outsider) is countered as a single line in the permutation of $COP^{'} \{ \alpha \} \{ \beta \}$. 
Shown in Figure 4 is the special case of one tree inside and one tree outside the circle. 
Complicated cases can be constructed in similar fashion.

\begin{figure}[h]
\begin{center}
\leavevmode
{\epsfxsize=4.0truein \epsfbox{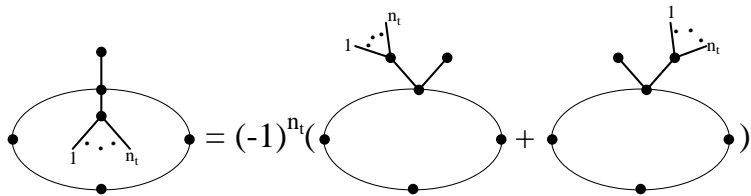}}
\end{center}
\caption{Reflecting a tree inside the circle to the outside. 
Note the order of the external lines after the reflection and the overall sign.}
\end{figure}

The second subtlety is that there are terms in $A_{n;c}$ constructed from $A_{n;1}$ which are proportional to
\be
\sum_{\sigma \in COP^{'}\{\alpha^{'}\}\{\beta^{'}\}}
{1 \ov \langle l, \sigma(1) \rangle}
\left(  \prod_{i=1}^{n-1} {1 \ov  \langle \sigma(i),\sigma(i+1) \rangle} \right)
{1 \ov \langle \sigma(n), l \rangle} 
\ee
where $\alpha^{'}/\beta^{'}$ indicate a subset of internal lines originally inside/outside of the circle,
with at least one set of them on a tree attached to the circle.
Graphically, it is shown in Figure 5.
Obviously one cannot obtain these term from the sub-leading MHV diagrams.
Fortunately these terms vanish upon summation over $\sigma$, thanks to the Schouten's identity again.

\begin{figure}[h]
\begin{center}
\leavevmode
{\epsfxsize=3.0truein \epsfbox{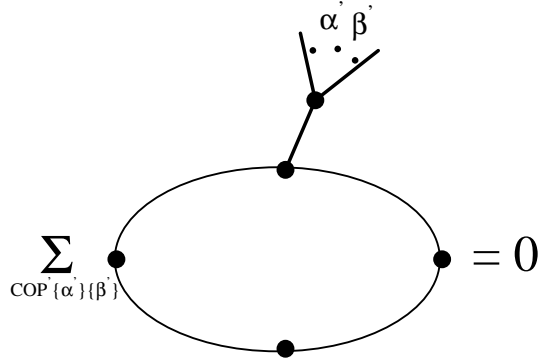}}
\end{center}
\caption{Graphic representation of Eq (4.7), 
indicating terms from $A_{n;c}$ constructed out of $A_{n;1}$ with the lines originally inside the circle
mingle with those originally outside.}
\end{figure}

With these two subtleties taken case of,
the rest of the proof follows exactly the same steps as those in the case of 1PI MHV diagrams.

\section{Conclusion}
In this paper, we have analyzed general amplitudes 
in general $N=4$ SYM theories to full one-loop in the paradigm of MHV Feynman diagrams. 
We classified all one-loop MHV Feynman diagrams.
It turns out there are only very limited number of integrals to be evaluated, 
though there are numerous loop diagrams.
For a process with $n$ external particles, there are only $[n/2]-1$ generically independent integrals. 
For a specific case with $q<[(n+1)/2]$ negative helicities,
only $q-1$ generically independent integrals need to be evaluated.
More so, the first $q-2$ integrals can be obtained 
from calculations for processes with lesser number of negative helicities.
For a specific case with $q>[(n+1)/2]$ negative helicities,
the result can be obtained from the case of $q^{'}=n-q$ via parity inversion.
These integrals depend crucially on $q$ but marginally on $n$.
All loop integrals can be obtained from these generic integrals by
appropriately substituting relevant helicities and momenta.
We have also prove that the relation between leading $N_c$ amplitudes $A_{n;1}$
and sub-leading amplitudes $A_{n;c}$ 
are identical to those obtained from conventional field theory calculations \cite{bern1}. 

As there are scant one-loop calculations beyond $q=2$ in the paradigm, 
these results are speculative to certain extent.
But given the strong evidence obtained so far, 
it is difficult to imagine the paradigm would go astray,
though an understanding of the approach from the perspective of conventional
quantum field theory is still wanting.
Actually, the relation between $A_{n;1}$ and $A_{n;c}$ as revealed in this paper 
could be interpreted as an indirect evidence to support the paradigm.
We suspect that the integrals will come out to be right.
It is now imperative to device an efficient way to calculate them.

\acknowledgments This work is supported in part by the National
Science Foundation of China.


\end{document}